\documentclass{iopart}
\usepackage{iopams} 
\usepackage{epsfig}
\usepackage{amssymb}

\usepackage{mathrsfs}
\usepackage{frcursive}
\usepackage{color}
\definecolor{darkblue}{rgb}{0,0,0.6}
\definecolor{darkred}{rgb}{0.6,0,0}
\usepackage{graphicx}

\usepackage[colorlinks=true,urlcolor=darkblue,citecolor=darkblue,linkcolor=darkred,hyperfootnotes=false]{hyperref}
\usepackage{url}

\newcommand{\cH}{\mathcal{H}}
\newcommand{\dd}{\textnormal{d}}

\newcommand{\ee}{\textnormal{e}}

\newcommand{\p}{\partial}

\newcommand{\eps}{\varepsilon}

\newcommand{\ellK}{\mathop{\textnormal{K}}}
\newcommand{\ellF}{\mathop{\textnormal{F}}}
\newcommand{\ellE}{\mathop{\textnormal{E}}}

\begin{document}

\title{Inactive dynamical phase of a  symmetric exclusion process on  a ring}

\author{Vivien Lecomte$^1$\footnote{\mailto{vivien.lecomte@univ-paris-diderot.fr}}, Juan P. Garrahan$^2$
and Fr\'ed\'eric van Wijland$^3$}

\address{$^{1}$Laboratoire de Probabilit\'es et Mod\`eles Al\'eatoires (CNRS UMR 7599), Universit\'e Paris Diderot -- Paris 7, Site Chevaleret, Case 7012, 75205 Paris cedex 13, France.\\
$^{2}$Department of Physics and Astronomy, University Park, Nottingham NG7 2RD, United Kingdom.\\
$^{3}$Laboratoire Mati\`ere et Syst\`emes Complexes (CNRS UMR
  7057), Universit\'e Paris Diderot -- Paris 7, 10 rue Alice Domon et L\'eonie
  Duquet, 75205 Paris cedex 13, France}

\begin{abstract}We investigate the nature of the dynamically inactive phase of a simple symmetric exclusion process on a ring. We find that as the system's activity is tuned to a lower-than-average value the particles progressively lump into a single cluster, thereby forming a kink in the density profile. All dynamical regimes, and their finite size range of validity, are explicitly determined. 
\end{abstract}
\maketitle

\section{Introduction}
The study of temporal large deviations in simple systems such as those of
interacting diffusing particles has revealed an unexpected wealth of
behaviors, ranging from puzzling correspondences between driven
systems and their equilibrium
counterparts~\cite{tailleurkurchanlecomte, tailleur_mapping_2008,
  imparato_equilibriumlike_2009, lecomteimparatovanwijland}, to
intrinsically dynamic phase transitions~\cite{bodineauderrida2,
  bodineauderrida3, garrahanjacklecomtepitardvanduijvendijkvanwijland,
  garrahan_first-order_2009,
  bertinidesolegabriellijonalasiniolandim-5,
  bertinidesolegabriellijonalasiniolandim-6}. The {\it dynamical phase
  transition} terminology dates back to the
eighties~\cite{beckschlogl}, in an era when the dynamical systems
community exploited the so-called thermodynamic formalism of Ruelle
and others to study and characterize the variety of dynamical regimes
displayed by simple iterated maps. There have been many efforts in the
recent past to classify the various phase transitions that can be
found between distinct dynamical regimes, but there is a great
scarcity of studies attempting an even coarse description of the
properties of these dynamical regimes. Exceptions are Bodineau and
Derrida's study of a weakly asymmetric exclusion
process~\cite{bodineauderrida2}, Toninelli and Bodineau's work on
kinetically constrained models~\cite{bodineautoninelli} and the finite
size study of Bodineau, Lecomte and
Toninelli~\cite{bodineaulecomtetoninelli}. In the generic dynamical
phase transition scenario, one of the dynamical phases is disordered
and its physics is easy to grasp, while the ordered symmetry-breaking
phase is usually hard to characterize, for in dynamics there is no
such thing as an easy-to-implement free-energy based variational
principle.

\medskip

In the present work, we would like to devote our own efforts to the
understanding of the nature of the various dynamical phases of the
Simple Symmetric Exclusion Process (SSEP) on a ring, a system of
mutually excluding particles hopping (if the target site is empty)
with equal rates to either of their nearest neighbor sites on a
one-dimensional lattice with periodic boundary conditions. The SSEP
belongs to the broader class of diffusive systems whose dynamics can
be described by fluctuating hydrodynamics, meaning that at the scale
fixed by the system size, the field of occupation numbers becomes a
smoothly varying function evolving through a Langevin equation with a
vanishingly small noise (as the system size increases). We choose to
classify the various time realizations the system can follow according
to their activity level defined as the number of particle moves having
taken place over a given time interval. In practice, however, we
introduce a Lagrange multiplier $s$ biasing trajectories towards a
given average activity~\cite{lecomte_thermodynamic_2007}. The $s>0$
(resp. $s<0$, $s\sim 0$) regime allows one to explore the
less-than-average-activity (resp. larger-than-average, typical)
trajectories. The latter activity is a physical observable that has
received considerable attention in the
past~\cite{garrahanjacklecomtepitardvanduijvendijkvanwijland,
  lecomte_thermodynamic_2007, merolle_spacetime_2005}, not only in
diffusive systems but also in a variety of systems with slow
dynamics~\cite{bodineau_lefevere_2008,hedges_dynamic_2009,pitard_dynamic_2011},
and which has strong connections with entropy production in
nonequilibrium systems~\cite{maesnetocnywynants}. This is the simplest
trajectory-dependent time-reversal symmetric observable one can think
of. For realizations showing typical or larger than typical activity,
the particles remain uniformly distributed, and the density profile
remains flat. This is the disordered regime. By contrast, if we focus
on realizations displaying a smaller than typical activity ($s<0$),
the uniform profile becomes unstable. This scenario was identified in
\cite{appertderridalecomtevanwijland}, and presents features similar to
the transition at high enough values of the current in the weakly
asymmetric exclusion process~\cite{bodineauderrida3}. The transition
point, occurring right at the typical (or average) activity is
characterized by a universal scaling function that does not depend on
the specifics of the SSEP, in the large size limit. The purpose of our
work is precisely to investigate the properties of the ordered phase
in which translation invariance, as our intuition tells us, will be
broken. The extreme low-activity behavior is indeed going to be
dominated by realizations in which particles have clustered into a
single lump, thus permitting particle moves at the borders of the
cluster, leading to a strongly subextensive activity.

\medskip

Before we proceed, we would like to recall that studying the
statistics of the activity in the SSEP is equivalent to investigating
the ground-state properties of a ferromagnetic $XXZ$
chain~\cite{appertderridalecomtevanwijland}, which allows one to
connect our work to the integrable systems literature. 
We are indeed interested in the smallest eigenvalue $-\psi(s)$
of the operator
\begin{equation}
 \mathbb H(s)= \frac L2 - \frac {\ee^{-s}}{2}\sum_{k=1}^L [\sigma_k^x\sigma_{k+1}^x+\sigma_k^y\sigma_{k+1}^y+\ee^s\sigma_k^z\sigma_{k+1}^z]
 \label{eq:defoperator}
\end{equation}
where the $\sigma^\alpha_i$'s are the Pauli matrices. In standard
notations (see Baxter~\cite{baxter} for instance) the anisotropy
parameter $\Delta$ is related to the Lagrange multiplier $s$ enforcing
a given average activity by $\Delta=\ee^s$. Another related
problem~\cite{liebliniger,gaudin} is that of determining the ground
state of a gas of hard-core bosons with interaction strength $-s$ and
density $\rho_0$, which, in the present $s>0$ case, would thus
correspond to attractive interactions (this is the no-saturation case
discarded by Lieb and Liniger at the beginning of their seminal
work~\cite{liebliniger}). In the regime of interest here, we have that
$\Delta>1$. The same problem
  of attractive bosons with hardcore interactions is of direct
  interest to study the directed polymer in the replica
  approach~\cite{kardar_replica_1987}, for which the problem was fully
  diagonalized~\cite{calabrese_free-energy_2010,dotsenko_bethe_2010},
  see~\cite{corwin_2011_arXiv:1106.1596} for a review. It was shown~\cite{albertinikorepinschadschneider} that
such a model and such a regime for $\Delta$ were relevant for the
description of the superconducting phase of some generalized Hubbard
models, when working at fixed magnetization. The latter
constraint, which renders the problem nontrivial (without the
constraint, see section 8.8 in Baxter~\cite{baxter}) turns out to be
our working framework, since particles are conserved in our closed
ring, and magnetization per site is $2\rho_0-1$. While the active
dynamical phase lent itself moderately easily to a Bethe Ansatz
approach~\cite{appertderridalecomtevanwijland}, which was shown to be
fully equivalent to the fluctuating hydrodynamics approach, it can be
understood from \cite{albertinikorepinschadschneider} that things only
get worse for $\Delta>1$. It therefore appears to be desirable to
resort an alternative method. It seems to us that the phase separation
mechanism conjectured in \cite{albertinikorepinschadschneider} has a
simple interpretation in our approach, as will become clear later in
the paper. Fluctuating hydrodynamics, in the integrable systems
language, can be viewed as the correct effective field theory able to
describe low lying excitations of the system.

\medskip

The questions we wish to answer can be phrased as follows. As the
system's activity is lowered below its average, how does the flat
density profile become unstable? Is there any limiting profile deep
into this inactive phase? Which are the activity scales, and what are
their system size dependence, governing these various dynamical
inactive regimes? 
Given that these are generic questions that can be raised for other comparable systems, it is of interest to present a model with a dynamic phase
transition for which they can be answered in an exact manner. 
We will begin with a brief overview of existing results
related to large deviations in the SSEP, as predicted by fluctuating
hydrodynamics---also termed Macroscopic Fluctuation Theory by Bertini
{\it et al.}~\cite{bertinidesolegabriellijonalasiniolandim-1,
  bertinidesolegabriellijonalasiniolandim-2,
  bertinidesolegabriellijonalasiniolandim-3,
  bertinidesolegabriellijonalasiniolandim-5,
  bertinidesolegabriellijonalasiniolandim-6}. This will allow us, in
Section \ref{sec:inactive}, to present explicit results on the density
profile in the inactive regime. Three scaling regimes exist, which we
will investigate separately in a mathematically controlled
fashion. Open questions will be gathered in the final section.

\section{The symmetric exclusion process and its fluctuating hydrodynamics description}\label{sec:overview}
\subsection{Fluctuating hydrodynamics}
We consider a Simple Symmetric Exclusion Process (SSEP) on an $L$-site
one-dimensional lattice with periodic boundary conditions. Each of the
mutually excluding $N$ particles can hop with rate 1 to either of its
nearest neighboring site, if empty. We consider time realizations of the
process over the time interval $[0,t]$ and we choose to work in space
and time units scaled by the system size $L$ and the diffusion
relaxation time $L^2$, respectively. We thus introduce a smoothly
varying field $\rho(x,\tau)$ of occupation numbers, defined by
$n_j(t')=\rho(j/L,t'/L^2)$, whose existence we assume. The field
$\rho$ evolves according to a conserving Langevin
equation\begin{equation}\label{fhyd} \p_\tau\rho+\p_x j=0
\end{equation}
with $j=-D\p_x\rho-\frac{1}{\sqrt{L}}\sqrt{\sigma}\xi$, and where
$D(\rho)=1$ and $\sigma(\rho)=2\rho(1-\rho)$ are functions reflecting
the SSEP dynamics at a macroscopic scale.  The Gaussian white
noise, $\xi(x,\tau)$, has unit correlations,
$\langle\xi(x,\tau)\xi(x',\tau')\rangle=\delta(x-x')\delta(\tau-\tau')$. The
Green-Kubo relation ensures that $\frac{2D}{\sigma}=\beta f''$, where
$f(\rho)$ is the free energy per unit length, which, in the SSEP,
reduces to the purely entropic expression $-\beta
f=-\rho\ln\rho-(1-\rho)\ln(1-\rho)$ [up to $\rho^2$, $f''$ is the
inverse isothermal compressibility, $\kappa_T=(\rho^2 f'')^{-1}$]. The
conditions under which fluctuating hydrodynamics is valid, and the
reasons why it applies to the SSEP, have been discussed by many
authors. We refer the reader to Kipnis and Landim's
book~\cite{kipnislandim} which specifically addresses the SSEP, and to
the series of papers by Bertini {\it et
  al.}~\cite{bertinidesolegabriellijonalasiniolandim-1,
  bertinidesolegabriellijonalasiniolandim-2,
  bertinidesolegabriellijonalasiniolandim-3,
  bertinidesolegabriellijonalasiniolandim-5,
  bertinidesolegabriellijonalasiniolandim-6}, the latter employing the
macroscopic fluctuation theory vocabulary. See
also~\cite{tailleur_mapping_2008} for a field theory approach using
coherent state construction of path integrals to represent the
$\langle \ee^{-sK}\rangle = \langle \ee^{-t \mathbb H(s)}\rangle$
[the operator $\mathbb H(s)$ is defined in~\eref{eq:defoperator}].

We will focus on the so-called activity $K(t)$, which, in the SSEP, is the number of particle hops that have taken place over a given time window $[0,t]$, over the whole ring. Up to irrelevant finite size corrections, we may view the activity as a functional of the local occupation numbers given by $K(t) =\int_0^d\dd t' \sum_{j,\pm} n_j(t') (1-n_{j\pm 1}(t'))$. In terms of the field of occupation
numbers, it is expressed by $K(t)=L^3\int_0^1\dd
x\int_0^{t/L^2}\dd\tau\sigma(\rho(x,\tau))$. The equilibrium
distribution in the SSEP is a simple Bernoulli distribution with
parameter $\rho_0$. Denoting by $\rho_0=N/L$ the average density, we
thus find that, to leading order in the system size, $\langle
K\rangle/(tL)=2\frac{N(L-N+1)}{L(L-1)}\simeq \sigma(\rho_0)$. For
time realizations with activity greater than $\langle K\rangle$ we
expect that the system remains as homogeneous as possible to favor
particle hops able to accommodate a high value of $K$.  We also
expect, as we confirm below, that for low activities the system will group particles
into a single cluster.  We now briefly recall what the existing
literature has established about these opposite regimes.

\subsection{Universal fluctuations of the activity}
The key to the calculation of large deviation properties in systems
described by a Langevin equation \eref{fhyd} is that, since the noise becomes
vanishingly small as the system size increases, a WKB-like saddle
point expansion can be exploited. There are many ways to implement the
saddle point expansion. One is to cast the generating function
$Z(s,t,L,\rho_0)=\langle\ee^{-s K}\rangle$ of the activity in the form
of a path-integral over a pair of fields $\bar{\rho}$ and $\rho$ (the
latter is the actual occupation number field, while the former is
conjugate to the noise $\xi$). The net result of that procedure that
has been described many times in the past (see
\emph{e.g.}~\cite{tailleur_mapping_2008}) is that
\begin{equation}
Z=\int{\mathscr D}\bar{\rho}{\mathscr D}\rho\,\ee^{-L S[\bar{\rho},\rho;s]}
\end{equation}
where the action $S[\bar{\rho},\rho;s]$ reads
\begin{equation}
S=\int_0^{t/L^2}\dd\tau\int_0^1\dd x\left[\bar{\rho}\p_\tau\rho+\p_x\bar{\rho}\p_x\rho-\frac{\sigma(\rho)}{2}(\p_x\bar{\rho})^2+sL^2 \sigma(\rho)\right]
\end{equation}
Once cast in the above path-integral formulation, it is clear that, as
$L\to\infty$, the leading behavior to $Z$ will be given by a saddle
point evaluation of the path-integral. One must then minimize the
action $S$ with respect to $\bar{\rho}$ and $\rho$. The latter gives
us the optimal density profile able to accommodate a given value of
the average activity, as tuned by $s$. In Appert {\it et
  al.}~\cite{appertderridalecomtevanwijland} it was shown that the
large deviation function of the activity, defined by
\begin{equation}
\psi(s)=\lim_{t\to\infty}\frac{\ln Z}{t}
\end{equation}
could be written in the form
\begin{equation}
\label{resultadlvw}
\psi(s)=-s\frac{\langle K\rangle}{t}+\frac{D}{L^2}{\mathcal F}\left(-\frac{\sigma\sigma''}{8 D}sL^2\right)
\end{equation}
where ${\mathcal F}$ is a scaling function independent of the functions
$D$ and $\sigma$, and its argument reduces to
${\mathcal F}\left(\frac{\sigma}{2}sL^2\right)$ in the SSEP. The functions $D$ and $\sigma$ appearing in this formula are evaluated at $\rho_0$, as in the sequel of the paper when no argument is made explicit. The
hypotheses underlying this result are that the optimal profile in the
path-integral formulation for the generating function $Z$ is both
stationary and uniform at the value $\rho_0$. The function
${\mathcal F}(u)$ exhibits a branch cut along the real axis for
$u\to(\pi^2/2)^-$, which signals that for $s>\frac{\lambda_c}{L^2}$,
$\lambda_c=\pi^2/\sigma(\rho_0)$, the uniform profile ceases to be the
one minimizing the action. The new results of the present work are
devoted to a study of this $sL^2>\lambda_c$ regime. We now define the parameter $\lambda$  by $\lambda=sL^2$.

It is piquant
to note that $\mathcal F$ has different integral representations
according to whether $\lambda<0$ or $0<\lambda<\lambda_c$. We need
it on the positive side, namely ${\mathcal F}(x)=2 x-4 \sqrt{2}
x^{3/2} \int_{-1}^1 \dd y\;y^2 \cot \left(\sqrt{2x} \sqrt{1-y^2}\right) $. The function $\mathcal F$ has the following
limiting behaviors
\begin{eqnarray}
{\mathcal F}(x)&\simeq \frac{2^{7/2}}{3\pi}(-x)^{3/2}\textnormal{ for }x\to-\infty\\
{\mathcal F}(x)&=C_1-C_2(\pi^2/2-x)^{1/2}+\ldots      \textnormal{ for }x\to\frac{\pi^{2}}{2}^-
\end{eqnarray}
where $C_1\simeq 51.61351...$ and $C_2\simeq 55.83091... >0$. The latter expansion explicitly displays the $\sqrt{\pi^2/2-x}$ singularity. The $x\to-\infty$ limiting behavior was first found by Lieb and Liniger in their study of the one-dimensional Bose gas with repulsive interactions~\cite{liebliniger}. We now turn to our analysis of the inactive phase.\\

\section{Inactive phase}\label{sec:inactive}

\subsection{The technical problem}
Inspired by the recent work of Bodineau and
Toninelli~\cite{bodineautoninelli}, we define a rescaled large
deviation function $\phi(\lambda)\equiv
\lim_{L\to\infty}L\lim_{t\to\infty}\frac{\ln\langle\ee^{-\lambda
    K/L^2}\rangle}{t}=\lim_{L\to\infty}L\psi(\lambda/L^2)$. From
\eref{resultadlvw} we get that $\phi(\lambda)=-\lambda\sigma(\rho_0)$ which has finite size corrections of the order $1/L$ given by
${\mathcal F}(\sigma\lambda/2)$. While ${\mathcal F}$ appears only
through finite-size corrections, we note that its singularities signal
the end of the regime over which the uniform saddle profile remains
the optimal one. Mathematically, the quadratic expansion of the action
around the uniform saddle ceases to be positive definite at
$\lambda>\lambda_c$. We therefore search for another saddle to the
Euler-Lagrange equations. Due to the conservation of the number of
particles, we introduce a Lagrange multiplier $\mu$ enforcing the
total density to be $\rho_0$, by adding a $\mu\int(\rho-\rho_0)$
contribution to the action.  Assuming a stationary saddle, the
Euler-Lagrange equations of motion read
\begin{equation}\label{el1}
j=-\p\rho+\sigma\p\bar{\rho},\;\;\p^2\bar{\rho}+\frac{\sigma'}{2}(\p\bar{\rho})^2-\lambda\sigma'-\mu=0
\end{equation}
We see a posteriori that in the homogeneous regime we must have
$\mu=-\lambda\sigma'(\rho_0)$. The only way to find out the stability
regime of the homogeneous profile is to study the quadratic action
expanded in the vicinity of that saddle. Since biasing the
trajectories with the activity does not break the left-right symmetry
nor the time-reversal one ($K$ is even by time-reversal), we search
for a solution to \eref{el1} in which the current $j=0$. Then it is
easy to verify that the equations in \eref{el1} are equivalent to
Euler-Lagrange equations deduced from a Lagrangian
$\mathcal{L}=\frac{(\p_x\rho)^2}{2\sigma}+\lambda\sigma+(\rho-\rho_0)\mu$,
and were the momentum conjugate to $\rho$ is $\pi=\frac{\p
  L}{\p\p_x\rho}=\frac{\p_x\rho}{\sigma}=\p_x\bar{\rho}$. The
corresponding Hamiltonian $\mathcal{H}=\pi \p_x\rho-\mathcal{L}$ is a
constant of motion, which means, when written in terms of the Lagrangian variables, that $\mathcal H$ given by
\begin{equation}
\mathcal{H}=\frac{1}{2\sigma}(\p\rho)^2-\lambda\sigma-(\rho-\rho_0)\mu
\end{equation}
is independent of $x$. A simple rewriting of the above equation tells us that
\begin{equation}\label{meca2}
\frac 12 (\p_x\rho)^2+E_P(\rho)=0,\; E_P(\rho)=-\lambda \sigma(\rho)^2-\big[(\rho-\rho_0)\mu+\mathcal{H}\big]\sigma(\rho)
\end{equation}
Again, \eref{meca2} has a simple mechanical interpretation: $\rho$ is
the position of a particle evolving in the force field given by the
potential energy $E_P(\rho)$, with kinetic energy
$\frac{1}{2}(\p_x\rho)^2$, while maintaining a zero total mechanical
energy. Time coordinate is $x$. The two constants $\mathcal H$ and
$\mu$ are for the moment undetermined. Periodic boundary conditions
and particle conservation will lead to $\lambda$ and $\rho_0$
dependent expressions for these parameters. In the homogeneous regime
($\lambda<\lambda_c$), we must have $E_P=0$, 
$\mathcal{H}=-\lambda\sigma(\rho_0)$ and
$\mu=-\lambda\sigma'(\rho_0)$. In the mechanical analogy, searching
for a periodic profile means that we are after a periodically
oscillating solution between two extreme positions. For
$\lambda<\lambda_c$ and the corresponding values of $\mathcal H$ and
$\mu$ there is only a single value of $\rho$ such that $E_P(\rho)=0$,
which means that the oscillations are reduced to a standstill at this
very density (which is of course $\rho_0$).
Note that, mathematically speaking, a similar set of equations were
obtained by Bodineau and Derrida~\cite{bodineau_current_2004} in the
context of the {\it additivity principle} for the large deviations of the
current in a system in contact with two reservoirs (a situation in which
particle number is not conserved and no chemical potential is required).
 We now set out to find the
optimal profile $\rho(x)$ along with $\mathcal H$ and $\mu$ in the
regime where $\lambda\to\lambda_c^+$.

\subsection{Close to the critical point}

We use this formulation to find the optimal profile for $\lambda=\lambda_c+\eps$, with $\eps\to 0^+$. In the spirit of \cite{bodineauderrida2}, we assume that a small perturbation around the flat profile will develop, which leads us to write that $\rho(x)=\rho_0+\sqrt{\eps}\phi_1(x)+\eps\phi_2(x)+\ldots$, $\mu=\mu_c+\sqrt{\eps}\mu_1+\eps\mu_2+\ldots$, with $\mu_c=-\pi^2\frac{\sigma'}{\sigma}$. We write that $\mathcal H$ is independent of $x$ order by order in $\eps$, which, after a few manipulations, leads to
\begin{equation}
\phi_1(x)=\frac{4}{\pi}[\rho_0(1-\rho_0)]^{3/2}\cos(2\pi x)
\end{equation}
and 
\begin{equation}
\phi_2(x)=-\frac{1}{2\pi^2}\sigma'(\rho_0)\sigma(\rho_0)^2\cos(4\pi x)
\end{equation}
together with $\mu_1=0$ and $\mu_2=-\sigma'(\rho_0)^3/4$. We then find that as $\lambda=\lambda_c+\eps$, with $\eps\to 0^+$,
\begin{equation}
\phi(\lambda)=-\pi^2-\sigma(\rho_0)\eps+3\frac{\sigma(\rho_0)^3}{\pi^2}\eps^2+\ldots
\end{equation}
We find that ${\mathcal H}=-\pi^2-\sigma(1-4\sigma)\eps+\ldots$. 
In other words, the instability sets in via the largest wave-length
Fourier mode, just as was the case in the system studied by Bodineau
and Derrida~\cite{bodineauderrida2}. However, in our case, we can go
deeper into the broken symmetry phase, as will now become clear.

\subsection{In the inactive regime}
\begin{figure}[tbp]
\begin{center}
\includegraphics[width=.75\columnwidth]{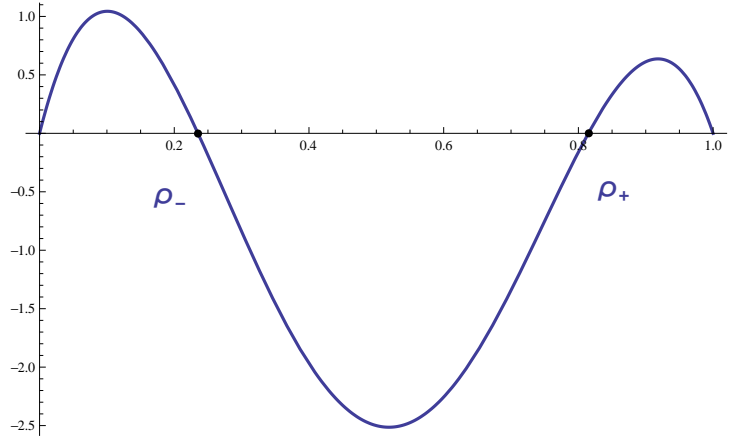}
\end{center}
\caption{The potential energy profile, $E_P(\rho)$, is shown for a set of typical values of the parameters $\lambda$, $H$, $\mu$ and $\rho_0$.
   \label{fig-Epot}}
\end{figure}

In order to pursue our mechanical analogy, we plot $E_P(\rho)$ as a
function of $\rho$ and have the space variable $x$ play the role of a
time. A typical plot is shown in figure \ref{fig-Epot}.
There one can see that $E_P(\rho)$ vanishes at two densities
$\rho_\pm$ given by
\begin{equation}
\rho_\pm=\frac{2\lambda+\mu\pm\sqrt{8\lambda({\mathcal H}-\mu\rho_0)+(2\lambda+\mu)^2}}{4\lambda}
\label{eq:rhopm}
\end{equation} 
The particle thus oscillates between these two extreme positions, and it takes a full period
\begin{equation}\label{const1}
\int_0^1\dd x=1=2\int_{\rho_-}^{\rho_+}{\dd\rho}\frac{1}{\sqrt{-2 E_P(\rho)}}
\end{equation}
to do the round-trip between $\rho_-$ and $\rho_+$. Besides, we must have that the total density is $\rho_0$, which in turn imposes
\begin{equation}\label{const2}
\rho_0=\int_0^1\dd x\rho(x)=2\int_{\rho_-}^{\rho_+}{\dd\rho}\frac{\rho}{\sqrt{-2 E_P(\rho)}}
\end{equation}
The two equations \eref{const1} together with \eref{const2} thus
completely fix the unknowns $\mathcal H$ and $\mu$. At this stage, it
is more convenient to deal with the $\rho_0=1/2$ case, for which
analytic expressions are somewhat simpler.

\subsubsection{At half-filling}

Particle-hole symmetry at $\rho_0=1/2$ imposes that
$E_P(\rho)=E_P(1-\rho)$, which in turn forces $\mu=0$ (this can
explicitly be verified from \eref{const1} and \eref{const2}). Then we
have that
\begin{equation}
0 <
\rho_-=\frac{\lambda-\sqrt{\lambda(2{\mathcal H}+\lambda)}}{2\lambda}
<
\rho_+=\frac{\lambda+\sqrt{\lambda(2{\mathcal H}+\lambda)}}{2\lambda}<1
\end{equation}
%
%
The density profile thus evolves between $\rho_-$ and $\rho_+$ given
above, with the constant $\mathcal H$ being given by the implicit
solution to
\begin{equation}\label{implicitH}
\frac 12
=\frac{(\sqrt{2\cH+\lambda}-\sqrt{\lambda})\ellK\left(\frac{2\sqrt{\lambda(2\cH+\lambda)}}{\cH+\lambda+\sqrt{\lambda(2\cH+\lambda)}}\right)}{\sqrt{2}\cH}
\end{equation}
where $\ellK$ is the complete elliptic integral of the first kind. Using that as $u\to 0$
\begin{equation}
\ellK (1-u)=2\ln 2-\frac 12 \ln u+\frac{u}{8}(-\ln u-2+4\ln 2)+{\mathcal O}(u^2)
\end{equation}
we can solve for $\cH$ as a function of $\lambda$ from \eref{implicitH}, which leads to
\begin{eqnarray}
  \cH=&8\lambda\ee^{-\sqrt{\frac{\lambda}{2}}}+
  16\ee^{-2\sqrt{\frac{\lambda}{2}}}\lambda\big(\sqrt{2\lambda}-4\big)\nonumber\\&\quad+
  16\ee^{-3\sqrt{\frac{\lambda}{2}}}\lambda(6\lambda-19\sqrt{2\lambda}+22)+
  {\mathcal O}\left(\lambda^{3/2}\ee^{-4\sqrt{\frac{\lambda}{2}}}\right)
\end{eqnarray}
The large deviation function $\phi(\lambda)$ is then obtained from
\begin{eqnarray}
\phi(\lambda)&=-\int_{0}^1\dd x\left[\frac{(\p_x\rho)^2}{2\sigma}+\lambda\sigma+(\rho-\rho_0)\mu\right]\\
&=\int_{0}^1\dd x\left[\cH-\frac{(\p_x\rho)^2}{\sigma}\right]
\label{bitphi}
\end{eqnarray}
Given that  we have 
\begin{equation} \label{bitphi2}\int_{0}^1\dd x\frac{(\p_x\rho)^2}{\sigma}=2\int_{\rho_-}^{\rho_+}\dd \rho\frac{\sqrt{-2E_P}}{\sigma}
\end{equation}
we arrive, after some manipulations, at
\begin{eqnarray}
\phi(\lambda)=-{\sqrt{2}} 
 \big(\sqrt{2\cH+\lambda}+\sqrt{\lambda}\big)\ellE\left(\frac{2\sqrt{\lambda(2\cH+\lambda)}}{\cH+\lambda+\sqrt{\lambda(2\cH+\lambda)}}\right)
\label{exactphihalf}
\end{eqnarray}
%
\begin{figure}[tbp]
\begin{center}
\includegraphics[width=.55\columnwidth]{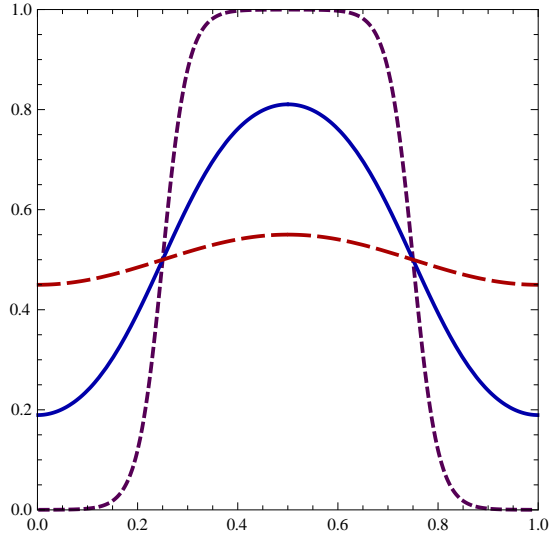}
\end{center}
\caption{The density profile is shown for $\lambda=\lambda_c+\frac
  1{10}$ (red, large dashes), $25$ (blue), $200$ (purple, small
  dashes) at average density $\rho_0=\frac 12$. The kink becomes steeper as
  $\lambda$ is increased. The slope of the rise to the plateau
  increases as $\sqrt{\lambda}$ for large $\lambda$.
   \label{fig-prof}}
\end{figure}
%
where $\cH$ as a function of $\lambda$ is extracted from the implicit formula \eref{implicitH}. The notations $\ellK$ and $\ellE$ refer to the complete elliptic integrals of the first and  second kind, respectively. This expression is valid for any $\lambda>\lambda_c$.
Using the following asymptotic formula for $E$ as $u\to 0$
\begin{eqnarray}
\ellE(1-u)&=1+u\left(-\frac 14 \ln u-\frac 14+\ln 2\right)\nonumber\\ &
\qquad +\frac{u^2}{64}\left(-6\ln u-13+24\ln 2\right)+{\mathcal O}(u^3)
\end{eqnarray}
we arrive at the following large $\lambda$ asymptotics for $\phi(\lambda)$:
\begin{eqnarray}
\phi(\lambda)=
&-2\sqrt{2\lambda}-8\sqrt{2\lambda}\ee^{-\sqrt{\frac{\lambda}{2}}} \nonumber\\
&\quad+8\big(2\lambda-3\sqrt{2\lambda}\big)
\ee^{-2\sqrt{\frac{\lambda}{2}}}+{\mathcal O}\left(\lambda^{\frac 32}\ee^{-3\sqrt{\frac{\lambda}{2}}}\right)
\end{eqnarray}
This is the final result for the $\rho_0=1/2$ particular
case. Translated into the language of quantum spin chains,
$\phi(\lambda)$ in \eref{exactphihalf} is the ground-state energy per
site of a ferromagnetic $XXZ$ chain with anisotropy parameter
$\Delta=1+\frac{\lambda}{L^2}$ for $\lambda>\lambda_c$, and at
half-filling. To our knowledge, this is the first appearance of this
expression in the literature.
The optimal profile
corresponding to that value of $\phi(\lambda)$ takes the form of a
smooth kink with area equal to $1/2$, as shown in figure
\ref{fig-prof} for increasing values of $\lambda$.
Its expression reads (for $x$ as function of $\rho$ for $0\leq x\leq \frac 12$)
\begin{eqnarray}
  \label{eq:profile_halffilling}
  \hspace*{-1cm}
  x(\rho) = \frac{1}{2}-
  \frac{\sqrt{2} \ellF\Big(\frac{1}{2} \arccos\left(\frac{(\rho-1)\lambda-H }{(1-\rho ) \sqrt{\lambda  (2 H+\lambda )}}\right)
    \Big|\frac{2 \sqrt{\lambda  (2 H+\lambda )}}{H+\lambda +\sqrt{\lambda  (2 H+\lambda )}}\Big)}
  {\sqrt{2 H+\lambda }+\sqrt{\lambda }}
\end{eqnarray}
where $\ellF$ is the incomplete elliptic integral of the first kind.

\subsubsection{At arbitrary density $\rho_0$}

\begin{figure}[tbp]
\begin{center}
\includegraphics[width=.55\columnwidth]{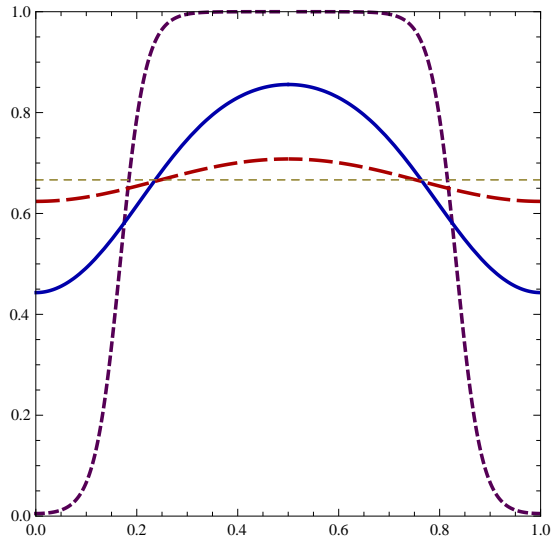}
\end{center}
\caption{The density profile is shown for $\lambda=\lambda_c+\frac
  1{10}$ (red, large dashes), $25$ (blue), $200$ (purple, small
  dashes) at average density $\rho_0=\frac 23$ (thin dashes). The kink becomes steeper as
  $\lambda$ is increased. 
   \label{fig-prof-rho23}}
\end{figure}

For general $\rho_0$, by exactly the same methods, though with
somewhat less elegant simplifications (no symmetry enforces the
Lagrange multiplier $\mu$ to vanish), the two equations \eref{const1}
and \eref{const2} now take the following forms
\begin{eqnarray}
  \label{const1-dur}
  \hspace{-2cm}
\frac12\rho_0=
\frac{2 \lambda  \ellK\Big(\frac{\mu +2\lambda  (1-2 \hat\rho )}{(\mu +2 \lambda  (1-\hat\rho)) (1-\hat\rho)}\Big)+
(\mu -2 \lambda  \hat\rho ) \Pi 
\Big(\frac{\mu +2\lambda  (1-2 \hat\rho )}{2 \lambda(1 - \hat\rho) }\Big|
\frac{\mu +2\lambda  (1-\hat\rho )}{(\mu +2 \lambda  (1-\hat\rho)) (1-\hat\rho)}\Big)}
{2 \lambda  \sqrt{(1-\hat\rho ) (\mu +2   \lambda  (1-\hat\rho ))}}
\end{eqnarray}
%
where $\Pi$ is the complete elliptic integral of the third kind, along with
\begin{equation}
\frac 12=\sqrt{\frac{1}{(1-\hat\rho ) (2 \lambda  (1-\hat\rho )+\mu )}} 
\ellK\left(\frac{\mu +2 \lambda  (1-2 \hat\rho )}{(\mu +2 \lambda  (1-\hat\rho )) (1-\hat\rho )}\right)
\label{const2-dur}
\end{equation}
%
%
\begin{figure}[ht]
\begin{center}
\includegraphics[width=.7\columnwidth]{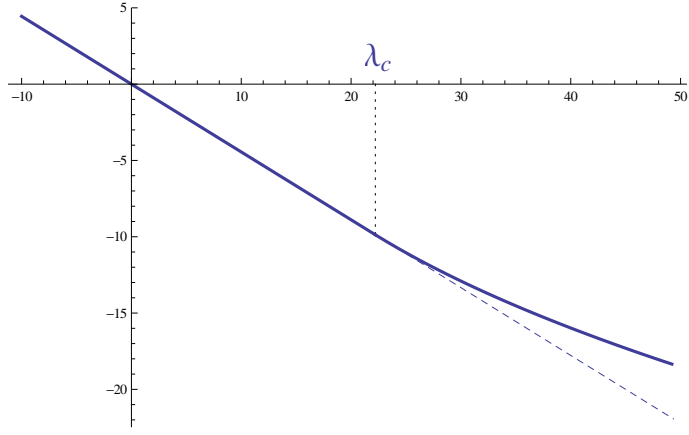}
\end{center}
\caption{Large deviation function $\phi(\lambda)$ (thick line) as a
  function of $\lambda$, at average density $\rho_0=\frac 23$.  The function is singular at
  $\lambda=\lambda_c=\frac{\pi^2}{\sigma(\rho_0)}$ (the second
  derivative is discontinuous), marking the entrance in the regime
  where the density profile is non-uniform ($\lambda>\lambda_c$).
  The value of the large deviation function for a uniform profile at $\lambda>\lambda_c$
  is shown for comparison (dashed line). 
    \label{fig:psi}}
\end{figure}
\begin{figure}[h]
\begin{center}
\includegraphics[width=.7\columnwidth]{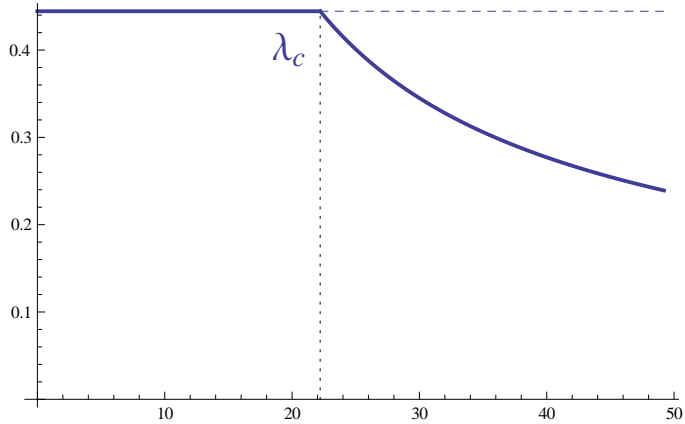}
\end{center}
\caption{Mean activity $\frac 1{Lt} \langle K\rangle_\lambda=-\phi'(\lambda)$ (thick blue line) as a
  function of $\lambda$, at average density $\rho_0=\frac 23$.  The function is singular at
  $\lambda=\lambda_c=\frac{\pi^2}{\sigma(\rho_0)}$.
  The value of $\frac 1{Lt} \langle K\rangle_s$ is shown for a uniform profile at $\lambda>\lambda_c$
  is shown for comparison (dashed line). As expected, above $\lambda_c$, the activity for histories 
  with a non-uniform profile is lower than for uniform profiles.
  \label{fig:psiprime}}
\end{figure}
%
where $\hat\rho$ is either equal to the minimum or the maximum
density $\rho_\pm$ [defined in~\eref{eq:rhopm}].
Once $\mu$ and $\cH$ have been extracted from \eref{const1-dur} and
\eref{const2-dur}, one can determine the large deviation function
$\phi(\lambda)$ as a function of $\lambda$ and $\rho_0$ from the
formula (see figures~\ref{fig:psi} and~\ref{fig:psiprime})
\begin{eqnarray}
\phi(\lambda)&=
-2 \sqrt{(1-\hat\rho ) (2 \lambda  (1-\hat\rho )+\mu )} 
\ellE\left(\frac{\mu +2 \lambda  (1-2 \hat\rho )}{(\mu +2 \lambda  (1-\hat\rho )) (1-\hat\rho )}\right)
\label{finalphi}
\end{eqnarray}
where $\lambda>\lambda_c$.%
%
%
The expansion close to $\lambda_c$ can be checked from those results. We see in particular
that $\frac 1{L^3t}\langle K^2\rangle_c=\phi''(\lambda)$ presents a jump $\phi''(\lambda_c^+)-\phi''(\lambda_c^-)=\frac{6\sigma(\rho_0)^3}{\pi^2}$.
We do not intend to labour beyond the leading order of $\phi(\lambda)$ as $\lambda\gg 1$, which is independent of $\rho_0$, 
\begin{equation}
\phi(\lambda)\simeq-2\sqrt{2\lambda}+\cH+(1-\rho_0)\mu+\ldots
\end{equation}
That the leading contribution is independent of $\rho_0$ could have been expected, as the main contribution to  $\phi$ as given in \eref{bitphi} is the integral \eref{bitphi2}, which is itself dominated by a small region of size $1/\sqrt{\lambda}$ where the derivative of $\rho$ varies the most steeply between $\rho_-$ and $\rho_+$, while density is close to $\frac{\rho_-+\rho_+}{2}\simeq 1/2$, and thus 
\begin{equation}
\phi(\lambda)\sim-\left({1}/{\sqrt{\lambda}}\right)\times\frac{1}{1/\sqrt{\lambda}^2}\sim-\sqrt{\lambda}
\end{equation}
Together with the implicit formulas \eref{const1-dur} and
\eref{const2-dur}, equation \eref{finalphi} is the ground state energy
of an $XXZ$ ferromagnetic Heisenberg chain with anisotropy parameter
$\Delta=1+\frac{\lambda}{L^2}$ at fixed magnetization $L(2\rho_0-1)$.
The density profile is, for $x$ as function of $\rho$ for $0\leq x\leq \frac 12$
(see figure~\ref{fig-prof-rho23}),
\begin{eqnarray}
  x(\rho)=\frac{
   \ellF\Big(\arcsin\Big(\sqrt{\frac{(\mu-2 \lambda \rho_- ) (\rho-\rho_- )}{(2(2 \rho_--1) \lambda -\mu ) (1-\rho )}}\Big)
     \Big|\frac{2\lambda(1-2 \rho_-) +\mu }{(\mu-2 \lambda \rho_- )\rho_-  }\Big)}
  {\sqrt{\rho_- (2 \lambda  \rho_--\mu )}}
\label{eq:profile_anyfilling}
\end{eqnarray}
%

\subsection{The infinitely inactive regime}
In the infinitely inactive regime, that is for $\lambda\gg L^2$, the
density profile described in the previous subsections becomes
infinitely steep and the assumptions underlying fluctuating
hydrodynamics fail. Since the typical slope of the rise of the profile
towards its plateau value increases at $\sqrt{\lambda}$, fluctuating
hydrodynamics must fail when the slope exceeds $L$, that is when
profile variations become significant at the lattice scale, which is
consistent with the $\lambda\gg L^2$ condition.  
For large $s$ it is easy to see that the ground state of the $s$-modified evolution operator (\ref{eq:defoperator}) reads
\begin{equation}
\psi(s)=-2L \sigma(\rho_0)(1-\ee^{-s}) + O(\ee^{-2s})
\end{equation}
The above regime consistently connects to the calculation of the
previous subsection where it was shown that
$\phi(\lambda)\sim-\sqrt{\lambda}$ for large values of $\lambda$. The
validity of that previous result holds until $\phi(\lambda)\sim
-2L\sigma(\rho_0)$, when it connects with the infinitely
inactive regime.  This occurs as $\lambda\sim L^2$ and confirms once
more that the crossover is at a typical $\lambda$ of the order of
$L^2$.

\section{Prospects}
The SSEP on a ring displays a phase transition from a homogeneous
state to a kink-like profile as the trajectories' activity is lowered
below its average value. We have described both the details of the
kink profile that develops and the large deviation function of the
activity itself in this nonuniform dynamical state. In the course of
our analysis, we have identified three distinct finite-size scaling
regimes.  The system dealt with in the present work exhibits a second
order continuous phase transition. It would certainly be quite
desirable to find similar solvable examples for first-order
transitions. The mechanism by which the symmetry breaking sets in may
be more subtle. This is work in progress.

\subsection*{Acknowledgements}
We would like to thank Thierry Giamarchi 
for interesting discussions on $XXZ$ related matters.  This work was supported in part by British Council France Alliance Project 09.013.

\bigskip

\section*{References}
\bibliography{ssep}{}
\bibliographystyle{plain_url}

\end{document}